\documentclass[aps,pra,floatfix,preprint,superscriptaddress,nofootinbib]{revtex4-1}
\usepackage[utf8]{inputenc}
\usepackage{amsmath}
\usepackage{amsfonts}
\usepackage{amssymb}
\usepackage{graphicx,color}
\usepackage{hyperref}
\usepackage{natbib}

\usepackage{tikz}
 \usetikzlibrary{decorations.pathmorphing,decorations.markings,trees,positioning,arrows}
 \usetikzlibrary{calc,patterns}
 \usetikzlibrary{shapes,backgrounds}   
\tikzset{
    photon/.style={decorate, decoration={snake}, draw=black},
    particle/.style={draw=black,postaction={decorate,
        decoration={markings,mark=at position 0.65 with {\arrow[draw=black]{>}}}}},
    antiparticle/.style={draw=black,postaction={decorate,
        decoration={markings,mark=at position 0.5 with {\arrow[draw=black]{<}}}}},
    gluon/.style={decorate, draw=orange,very thick, 
	    decoration={coil,amplitude=4pt, segment length=6pt}},
    higgs/.style={draw=red,very thick, postaction={decorate},
	           decoration={markings,mark=at position .55 with  {\arrow[draw=red]{>}}}},
     doubleline/.style={draw=black,very thick,
            double distance=8pt,
            postaction={decorate,
            decoration={
                markings,
                mark=between positions 10pt and -10pt step 50pt with {
                    \arrow[very thick,yshift= 4pt,xshift=.8pt]{>}
                    \arrow[very thick,yshift=-4pt,xshift=.8pt]{<}
                },
            }},
        }}
\usetikzlibrary{shapes.misc}
\tikzset{cross/.style={cross out, draw=black, minimum size=2*(#1-\pgflinewidth), inner sep=0pt, outer sep=0pt},
cross/.default={1pt}}

\hypersetup{
colorlinks=true,
linkcolor    = blue,
citecolor   = blue,
    pdfborder={0 0 0},
}

\newcommand{\s}{\sigma}
\newcommand{\g}{\gamma}
\newcommand{\oo}{\mathcal{O}}
\numberwithin{equation}{section}

\begin{document}
\count\footins = 1000
\title{Thermalization/Relaxation in integrable and free field theories:\\
an Operator Thermalization Hypothesis}
\author{Philippe Sabella-Garnier}
\author{Koenraad Schalm}
\author{Tereza Vakhtel}
\affiliation{Instituut-Lorentz, $\Delta$ITP, Universiteit Leiden, P.O. Box 9506, 2300 RA Leiden, The Netherlands}
\author{Jan Zaanen}
\affiliation{Instituut-Lorentz, $\Delta$ITP, Universiteit Leiden, P.O. Box 9506, 2300 RA Leiden, The Netherlands}
\affiliation{Department of Physics, Stanford University, Stanford CA 94305, USA}

\begin{abstract}
Free or integrable theories are usually considered to be too constrained to thermalize. For example, the retarded two-point function of a free field, even in a thermal state, does not decay to zero at long times. On the other hand, the magnetic susceptibility of the critical transverse field Ising is known to thermalize, even though that theory can be mapped by a Jordan-Wigner transformation to that of free fermions. We reconcile these two statements by clarifying under which conditions conserved charges can prevent relaxation at the level of linear response and how such obstruction can be overcome. In particular, we give a necessary condition for the decay of retarded Green's functions. We give explicit examples of composite operators in free theories that nevertheless satisfy that condition and therefore do thermalize. We call this phenomenon the \emph{Operator Thermalization Hypothesis} as a converse to the Eigenstate Thermalization Hypothesis. 
\end{abstract}
\maketitle

\tableofcontents

\section{Introduction}
There has been a profound realization recently that the non-equilibrium dynamics of integrable systems is much richer than expected. Following the principle of entropy maximization, it is well known that a system with a conserved charge equilibrates/thermalizes to the Gibbs ensemble $\rho = \exp(-\beta(H-\mu Q))$. By extension a system with an infinite set of conserved charges ought then equilibrate to a Generalized Gibbs Ensemble (GGE) \cite{2007PhRvL..98e0405R,Caux:2013ra,2015PhRvL.115o7201I,2016JSMTE..06.4006C,2016JSMTE..06.4007V,2016JSMTE..06.4002E,2017PhRvB..95k5128I} $\rho_{GGE} = \exp(-\beta(H-\sum_i\mu_i Q_i))$. On the other hand, it is a conventional wisdom in physics that free or integrable theories do not thermalize. In classical mechanics this statement is exact. In a theory which has as many conserved charges as dynamical degrees of freedom, one transforms to action-angle variables. In these variables the motion can be solved independent of the initial conditions: the momenta are constant and the positions are linear functions in time. The system therefore never ``forgets'' its initial conditions, even with small uncertainties, and never equilibrates or thermalizes.

Both theoretical and experimental results in the recent decade have revealed that in quantum integrable theories it is the former which happens. After a quench or a (strong) disturbance of the system, it {\em equilibrates} to the GGE. There is dissipation and thermalization and some information is lost \cite{2016JSMTE..06.4007V,2016JSMTE..06.4002E,Castro-Alvaredo:2016cdj,Bertini:2016tmj,2017arXiv171100873C,2017PhRvB..95k5128I,2018PhRvL.120d5301D,DeNardis:2018omc}. Clearly, in quantum theories whether a system ``thermalizes'' is more subtle.\footnote{The quick argument for this is that in closed quantum systems in particular, unitarity prevents the evolution of any pure state into the mixed state given by the thermal density matrix. Quantum thermalization is therefore usually understood as a process of dephasing: a generic state is a superposition of energy eigenstates, each of which picks up a different phase under unitary evolution by the Hamiltonian. If the distribution of energies is dense enough, then in the thermodynamic limit it is possible to have an exponential decay of two-point functions. An easy way to see this is to ask what the recurrence time is. The phase of each energy eigenstate becomes one when the time is an integer multiple of $2\pi/E$, where $E$ is the energy of the state. Therefore, if the time elapsed is $2\pi$ times the lowest common multiple of all the energies involved then the system will return to its initial state. In the ``thermodynamic'' limit, i.e. with an infinite number of non-commensurate frequencies, the recurrence time is infinite. This is the physics behind the Eigenstate Thermalization Hypothesis (ETH) \cite{PhysRevA.43.2046,1994PhRvE..50..888S,Yukalov:2012pi,DAlessio:2016rwt}. However, the conventional ETH is formulated for theories with no conserved charges (although see \cite{Dymarsky:2018lhf,Maloney:2018yrz,Dymarsky:2019etq} for recent work to extend this). Moreover, we shall consider perturbations around a mixed state state, arguably equivalent to an open quantum system coupled to a heat bath.}
This finding of GGE formation is in tension with the intuition from classical mechanics. It could be assumed that integrable (quantum) field theories do not thermalize either since the number of conserved quantities constrains the allowed dynamics too much. The simplest examples are free field theories: with no scattering, the system would not thermalize and any perturbation would not decay but rather persist forever.

It is appropriate at this point to be explicit about what we shall mean by thermalization. For the purposes of this paper, we will consider a minimalist scenario where we perturb an otherwise given thermal state, and study the linear response. 
Mathematically, this perturbation can be captured by changing the Hamiltonian infinitesimally
\begin{equation}
\delta H=\int d^{d-1}x \phi(x,t) \oo(x,t)~,
\end{equation}
where $\phi(x,t)$ is the source profile, exciting the (potentially composite) Hermitian operator $\oo$ built out of the dynamical degrees of freedom of the system. 
The retarded two-point function of $\oo$ with respect to the state of the system --- in our case the thermal state --- then gives the leading response of the expectation value of the operator: 
\begin{align}
\delta \langle \oo(x,t) \rangle = \int d^{d-1}x' dt' G_R(x-x',t-t') \phi(x',t')~ \nonumber\\
G_R(x-x',t-t')=-i\Theta(t-t')\langle [\oo(x,t),\oo(x',t')]\rangle_\beta~,
\end{align}
where by $[\cdot,\cdot]$ we mean whichever bracket is appropriate for the spin-statistics of $\oo$.
If this retarded thermal two-point function of the operator dual to this perturbation exhibits an exponential decay at long times, then we will say thermalization occurs. To be precise we will say that {\em this perturbation thermalizes.} Because exponential decay means that the initial perturbation relaxes away, its initial condition is ``forgotten'' and the response to the perturbation is lost at long times. This ``forgetting initial conditions'' we shall use is a somewhat broader definition than the more specific one where one demands that the system ---as measured through one-point functions of local observables --- approaches a Boltzmann or (generalized) Gibbs ensemble as the final state, i.e. entropy maximization. 

One might object that thermality is already wired into the problem, so to speak, as we consider a thermal mixed state. This is not so, as one can easily reason. Even in a classical integrable theory one can build a thermal ensemble by hand. A perturbation thereof, however, should be infinitely long-lived by the same reasoning as before.  This can be computed explicitly in the simplest possible example, that of a (real) free scalar field, where the converved charges are the occupation numbers of each separate momentum mode. The Euclidean Green's function of the scalar field itself obeys the following equation:
\begin{equation}
\left(-\partial_\tau^2-\nabla^2+m^2\right)G_E(\vec{x},\tau)=\delta(\vec{x},\tau)~.
\end{equation}
The solution is easily found by going to momentum space. Imposing periodicity in $\tau$ to account for finite temperature leads to
\begin{equation}
G_E(\omega_n, \vec{k})=\frac{1}{\omega_n^2+\vec{k}^2+m^2}~,
\end{equation}
where $\omega_n$ are the Matsubara frequencies, $\omega_n=2\pi T n$, with $T$ the temperature and $n\in \mathbb{N}$. The retarded Green's function is obtained from the Euclidean one with the following prescription:
\begin{equation}
G_R(\omega,\vec{k})=- G_E(-i\omega+\eta,\vec{k})~,
\end{equation}
where $\eta$ is a positive infinitesimal term chosen to ensure that all the poles are below the real axis. Thus
\begin{equation}
G_R(\omega,\vec{k})=\frac{1}{(\omega+i\eta)^2-\vec{k}^2-m^2}~. \label{eq:gr}
\end{equation}
Fourier transforming back to position space, we obtain
\begin{align}
G_R(t,\vec{k})&=\frac{1}{2\pi}\int_{-\infty}^\infty d\omega \frac{e^{-i\omega t}}{(\omega+i\eta)^2-\vec{k}^2-m^2} \nonumber \\
&=-\frac{1}{\sqrt{\vec{k}^2+m^2}}\sin\left(\sqrt{\vec{k}^2+m^2}~t\right) \Theta(t)~. \label{eq:mixed-prop}
\end{align}
where we may set $\eta=0$ after it has served its purpose to shift the poles slightly below the real $\omega$ axis, thus ensuring that the retarded Green's function vanishes for $t<0$. Setting the regulator $\eta=0$, however, places the pole on the real line from below and this means that for $t>0$ the Green's function does not decay. Thus a direct perturbation of any of the degrees of freedom in a free theory does not relax away, exactly following the reasoning espoused above.\footnote{In many integrable, but not necessarily free theories, two-point functions of operators which are charged under a number of charges do not decay as can be shown by the Mazur bound. These {\em operators} as opposed to the system are called non-ergodic; see e.g. \cite{2013CMaPh.318..809I}.}
  
In order for the retarded Green's function to decay in time, its poles in the complex $\omega$ plane must be situated a finite distance below the real axis. For example, in an interacting theory $m^2$ is replaced with the self-energy, which can have an imaginary part.

\bigskip

It appears we have just confirmed the conventional wisdom that integrable theories do not thermalize. Consider, however, another example: the transverse field Ising model in 1+1 dimensions. It has the following Hamiltonian:
\begin{equation}
H=-J\sum_{i=1}^N \left(g\sigma_i^x + \sigma_i^z \sigma_{i+1}^z\right)~,
\end{equation}
with $[\sigma_i^a,\sigma_j^b]=2i\epsilon_{abc} \delta_{ij}$.
From a seminal result by Damle and Sachdev, it is known that, at the critical point ($g=1$) in the continuum limit at high temperatures, the retarded correlator for the $\sigma^z$ operator takes the following form \cite{Damle:1997rxu,sachdev_2011}:
\begin{align}
G_R^{(\sigma_z)} (\omega,k) \sim \frac{1}{T^{7/4}} \frac{\Gamma(7/8)}{\Gamma(1/8)} \frac{\Gamma\left(\frac{1}{16}-i\frac{\omega+k}{4\pi T}\right)\Gamma\left(\frac{1}{16}-i\frac{\omega-k}{4\pi T}\right)}{\Gamma\left(\frac{15}{16}-i\frac{\omega+k}{4\pi T}\right)\Gamma\left(\frac{15}{16}-i\frac{\omega+k}{4\pi T}\right)}~.
\label{eq:BTZ}
\end{align}
This correlation function has two infinite lines of poles, corresponding to when the argument of one of the gamma functions in the numerator is zero or a negative integer: \begin{equation}
\omega=-4\pi i T\left(n+\frac{1}{16}\right) \pm k~~,~~n=0,1,2,\cdots
\end{equation}
The $\sigma_z$ perturbation thus thermalizes away. This has ever since been the prototypical example of the crossover from the {\em coherent collisionless} to {\em incoherent hydrodynamic} regime for $\omega < T$ near a quantum critical point \cite{Damle:1997rxu,sachdev_2011} and Planckian dissipation \cite{2004Natur.430..512Z}. It is well known, however, that
the transverse field Ising model at the critical point is an integrable theory in disguise. After a change of variables it can in fact be mapped to a {\em free} fermionic theory. The relaxation and dissipation of the operator $\sigma_z$ is thus {\em in direct conflict} with the conventional wisdom that free field theories do not thermalize. What is going on?

Counter to the conventional wisdom, we will show that even free field theories, and by extension integrable theories, thermalize in the sense that most perturbations of the system (around the thermal state) {\em will} relax away. The proof of this result will in fact be rather elementary. We will make precise the intuition that integrable and free field theories are constrained and therefore not {\em all} perturbations in such theories can relax. Nevertheless there are always perturbations that {\em do} relax. Once one has understood how, it will most likely be as obvious that integrable systems thermalize, as the converse appeared to be before.  

This does leave the pointed question: how then does the existence of an infinite set of conserved charges affect the process of thermalization of a perturbation (as we've defined it above)?
It would seem natural that more conservation laws makes it harder for a perturbation to thermalize, but it cannot make it impossible, as illustrated by the example of the transverse field Ising model. 

We posit as the answer to this question a simple no-go condition for thermalization.
For this, consider the retarded Green's function in the thermal state:
\begin{align}
G_R(x,t)&=-i\Theta(t)\langle [\oo(x,t),\oo(0,0)]\rangle_\beta \nonumber\\
&=\frac{-i\Theta(t)}{Z(\beta)} \sum_n e^{-\beta E_n} \left[\langle n|\oo (x,t) \oo(0,0)|n\rangle - (-1)^{2s}\langle n|\oo(0,0) \oo(x,t)|n\rangle\right]~.
\end{align}
Here $s$ is the spin of the operator (which is continuous in 2D and either half-integer or integer in higher dimensions) and
\begin{equation}
\oo (x,t)=e^{iHt} e^{-i\vec{P}\cdot \vec{x}} \oo(0,0)e^{-iHt} e^{i\vec{P}\cdot \vec{x}}~.
\end{equation}
Inserting a complete set of states,  we can write
\begin{align}
G_R(\vec{x},t)=\frac{-i\Theta(t)}{Z(\beta)} \sum_{m,n} e^{-\beta E_n} \left[e^{-i(E_m-E_n)t+i(\vec{P}_m-\vec{P}_n)\cdot \vec{x}} - (-1)^{2s} e^{i(E_m-E_n)t-i(\vec{P}_m-\vec{P}_n)\cdot \vec{x}}\right] |\langle m|\oo|n\rangle|^2~,
\end{align}
and transforming to momentum space, we obtain:
\begin{align}
G_R(\vec{k},\omega)=-\frac{i}{Z(\beta)}\sum_{n,m} e^{-\beta E_n}\left[ \frac{\delta^{(d-1)}\left(\vec{k}-(\vec{P}_n-\vec{P}_m)\right)}{\omega+(E_n-E_m)+i\eta} - (-1)^{
2s}\frac{\delta^{(d-1)}\left(\vec{k}-(\vec{P}_m-\vec{P}_n)\right)}{\omega+(E_m-E_n)+i\eta}\right] |\langle m|\oo|n\rangle|^2~.
\end{align}
The asymmetry between $\vec{k}$ and $\omega$ is due to the step function, and $\eta$ is an infinitesimal positive term. 

We now state the no-go condition:
\begin{itemize}
\item[] If the non-zero values of $|\langle m|\oo|n\rangle|^2$ are such that fixing $\vec{P_n}-\vec{P_m}$ automatically fixes $E_n-E_m$ (or restricts it to a finite number of values), i.e. if $|\langle m|\oo|n\rangle|^2=0$ unless $E_n-E_m=F^{(\oo)}_{i}(P_n-P_m)$, where $F^{(\oo)}_{i}(P)$ are (not-necessarily continuous) functions that depend on $\oo$, then the perturbation $\oo$ will not thermalize. The number of such functions, $N$, must be finite and independent of the system size.
\end{itemize}
The proof of this condition is straightforward. If it holds, 
then we can write
\begin{align}
G_R(\vec{k},\omega)=\sum_{i=1}^{N}\left(\frac{H^{(\oo)}_i(\beta,\vec{k})}{\omega+F_i^{(\oo)}(\vec{k})+i\eta} - (-1)^{2s}\frac{H^{(\oo)}_i(\beta,-\vec{k})}{\omega-F_i^{(\oo)}(-\vec{k})+i\eta}\right) 
\end{align} 
where 
\begin{equation}
H^{(\oo)}_i(\beta,\vec{k})=-\frac{i}{Z(\beta)}\sum_{\substack{m,n|\\E_n-E_m\\=F_i^{(\oo)}(P_n-P_m)}} e^{-\beta E_n}\delta^{(d-1)}(\vec{k}-(\vec{P}_n-\vec{P}_m))  |\langle m|\oo|n\rangle|^2~,
\end{equation}
becomes a function that is independent of $\omega$.
Noting also that $F_i^{(\oo)}$ must be real by definition, the retarded Green's function therefore manifestly only has poles on the real $\omega$ axis, and there can be no thermalization.

We conjecture that the converse is also true. If this condition is violated, i.e. if the operator $\oo$ is such that $|\langle m|\oo|n\rangle|^2$ can depend on $E_n-E_m$ and $P_n-P_m$ independently, then it will thermalize. We call this the {\em Operator thermalization hypothesis} (OTH), by analogy with the well-known eigenstate thermalization hypothesis (ETH). In the case of ETH, thermalization occurs because the spectrum is chaotic enough to allow a full exploration of the Hilbert space with even a simple operator. OTH is in some sense the mirror image: the spectrum is very organized, but the operator is complex enough to cause that same exploration. 

The point of this article is that even in free theories one can find operators that violate the no-go condition. Our no-go condition/OTH can often be directly applied to most integrable theories in terms of their quasi-momentum. Clearly the presence of an infinite set of conserved charges constrains the dynamics severely, and restricts the choice of operators that will satisfy the OTH. Nevertheless this set is never empty.

The rest of this paper is devoted to an examination of various situations where thermalization occurs or does not occur (for similar observations in other contexts, see e.g \cite{Grozdanov:2015nea,Amado:2017kgr, Parker:2018yvk,2019arXiv190508266M, Banerjee:2019iwd,Craps:2019pma}) and the role played by this no-go theorem in each of these cases. In section 2, we consider two-dimensional conformal field theories. 2D CFTs are constrained enough that their thermal correlation functions are universal. Therefore, the simple example of a two-dimensional free field, which we treat explicitly, can teach us a lot as it is also a 2D CFT. In section 3, we revisit the transverse field Ising model and see how the results we've outlined above can be understood by mapping it to a 2D CFT. In section 4, we consider in more detail higher-dimensional free fields to show that our conclusions are not simply a consequence of the constrained nature of kinematics in two dimensions. Finally, we offer some concluding remarks in section 5.

\section{2D CFT}

Two-dimensional (relativistic) CFTs are probably the most well-known examples of theories with an infinite set of conserved charges aside from free fields. The conserved charges are encoded in the full Virasoro algebra; in essence they are integrals of polynomials of the stress tensor and its derivatives. A mutually commuting set of these can be built using the KdV hierarchy \cite{Bazhanov:1994ft}. We will first discuss 2D CFTs in general, and then specialize to the 2D free massless bosonic $c=1$ theory.

In 2D CFTs (on a line, i.e. a spatial coordinate of infinite extent), the finite-temperature Euclidean two-point function of primary operators is uniquely fixed by conformal invariance to be:
\begin{equation}
\langle T_E\left\{\oo(x,\tau) \oo(0,0)\right\} \rangle_\beta = \left(\frac{2\pi^2}{\beta^2}\right)^{h+\bar{h}} \frac{1}{\sinh^{2h}\left(\frac{\pi}{\beta}(x+i\tau)\right)\sinh^{2\bar{h}}\left(\frac{\pi}{\beta}(x-i\tau)\right)}~,
\end{equation}
where $h$ and $\bar{h}$ are, respectively, the holomorphic and anti-holomorphic weight of the operator $\oo$, which has scaling dimension $\Delta=h+\bar{h}$ and spin $s=h-\bar{h}$. $T_E$ denotes Euclidean time ordering:
\begin{equation}
\langle T_E\left\{\oo(x,\tau) \oo(0,0)\right\} \rangle_\beta=\langle \oo(x,\tau) \oo(0,0) \rangle_\beta \Theta(\tau)+ e^{2\pi i(h-\bar{h})}\langle \oo(0,0) \oo(x,\tau) \rangle_\beta \Theta(-\tau)~.
\end{equation}
The pre-factor in front of the second term accounts for spin-statistics in two dimensions; it equals $+1$ when $\oo$ has integer spin and $-1$ when it has half-integer spin, as expected for bosons and fermions.  We can obtain real-time correlators by analytically continuing:
\begin{equation}
\tau\rightarrow it \pm \epsilon~,
\end{equation}
picking the sign of the real infinitesimal part of $\tau$ to obtain the desired ordering. With the above prescription, one obtains the retarded Green's function 
\begin{align}
G_R(x,t)&=-i\Theta(t)\left(\langle \oo(x,t) \oo(0,0)\rangle_\beta - e^{2\pi i(h-\bar{h})} \langle \oo(0,0) \oo(x,t) \rangle_\beta\right) \nonumber \\
&=-i\Theta(t) \left(\frac{2\pi^2}{\beta^2}\right)^{h+\bar{h}} \left[
\frac{1}{\sinh^{2h}\left(\frac{\pi}{\beta}(x-t+i\epsilon)\right)\sinh^{2\bar{h}}\left(\frac{\pi}{\beta}(x+t-i\epsilon)\right)}\right. \nonumber \\
&-\left. \frac{1}{\sinh^{2h}\left(\frac{\pi}{\beta}(x-t-i\epsilon)\right)\sinh^{2\bar{h}}\left(\frac{\pi}{\beta}(x+t+i\epsilon)\right)} \right]
\end{align}
If $x-t$ and $x+t$ are {\em both} non-zero, this vanishes. Causality always implies that the retarded Green's function vanishes for $|x|>|t|$. The fact that it also vanishes for $|x|<|t|$ is a consequence of conformal field theories describing massless excitations; such excitations behave exactly as a free massless field with support on the light-cone only. This appears to severely restrict the possibility of long-time exponential decay. However, pick $x=t$ without loss of generality. Then one obtains
\begin{equation}
G_R(x=t,t)\sim \frac{1}{\sinh^{2\bar{h}}\left(\frac{2\pi t}{\beta}\right)}~.
\end{equation}
where we have suppressed the infinite pre-factor, which is constant in time, to better highlight the long-time exponential decay. Thus a generic operator in a 2D CFT {\em does} thermalize/relax, but essentially as it if it lives only on the 1D lightcone. We can think of the information from the perturbation as now spreading along both ``branches'' of the future light cone. It relaxes ballistically rather than diffusively \cite{2017NatSR...7.6118R}. Therefore, even if we follow one of the lightcones (say $x=t$), it can still decay and some information is lost. 
We will show below that this is a typical example of our Operator Thermalization Hypothesis.

To have {\em no} operator thermalization one needs further constraints. 
Consider a chiral primary operator, i.e. one with $\bar{h}=0$. Its retarded Green's function is fixed to be
\begin{align}
G_R(x,t)=-i\Theta(t)\left(\frac{2\pi^2}{\beta^2}\right)^h \left[
\frac{1}{\sinh^{2h}\left(\frac{\pi}{\beta}(x-t+i\epsilon)\right)} - \frac{1}{\sinh^{2h}\left(\frac{\pi}{\beta}(x-t-i\epsilon)\right)}\right]~. \label{eq:gx_chiral}
\end{align}
Given that $\epsilon$ is infinitesimal, this vanishes for any $x\neq t$. For $\epsilon \ll |x-t| \ll 1$, we can write:
\begin{equation}
G_R(x,t)=-i\Theta(t) 2^h \left[\frac{1}{(x-t-i\epsilon)^{2h}}-\frac{1}{(x-t+i\epsilon)^{2h}}\right] 
\end{equation}
This does not decay in time: it is either infinite (on the light cone, independent of time) or zero (off the light cone). To get a better feel for this, consider the special case where $2h$ is an integer. Note that 
\begin{equation}
\frac{1}{2\pi i} \lim_{\epsilon\rightarrow 0} \left[\frac{1}{u-i\epsilon} - \frac{1}{u+i\epsilon}\right]=\delta(u)~~,
\end{equation}
and so, for any integer $n>1$:
\begin{equation}
\lim_{\epsilon \rightarrow 0}\left[\frac{1}{(u-i\epsilon)^n}-\frac{1}{(u+i\epsilon)^n}\right]=(-1)^n \frac{2\pi i}{(n-1)!} \frac{d^{n-1}}{du^{n-1}} \delta(u)~,
\end{equation}
(understood as a distribution). 
Physically, for any time $t$ there is a position $x$ so that the perturbation to the system can be recovered. 

These pictures of real-time decay are fairly intuitive but the subtleties of dealing with the $i\epsilon$ prescription clearly play a role. We can be more precise by going to frequency space. A treatment in full generality would be tedious, so we limit our attention to two relatively simple cases that illustrate our point. A useful identity is \cite{,Gubser:1997cm,Bredberg:2009pv}:
\begin{equation}
\int dx e^{ikx} \left(\frac{\pi}{\beta\sinh\left[\frac{\pi x \pm i\epsilon}{\beta}\right]}\right)^{2h}=e^{-i\pi h} \left(\frac{2\pi}{\beta}\right)^{2h-1} e^{\mp \frac{k}{2T}} \frac{|\Gamma\left(h+\frac{ik\beta}{2\pi}\right)|^2}{\Gamma(2h)} \equiv G_{\pm}(k)~.
\end{equation}
Defining $x_{\pm}=x \pm t$, $k_\pm=\frac{k\pm\omega}{2}$ and Fourier transforming the step function in the $x_{+}$ direction first, we obtain that the Fourier transform of the finite temperature chiral operator Green's function \eqref{eq:gx_chiral} is
\begin{equation}
G_R(k,\omega)=\frac{2^{h-1}}{\pi(k+\omega)}\left[G_{+}(k)-G_{-}(k)\right] - i 2^{h-2}\left[G_{+}\left(\frac{k-\omega}{2}\right)-G_{-}\left(\frac{k-\omega}{2}\right)\right] \delta\left(\frac{k+\omega}{2}\right)~.
\end{equation}
We see that the only pole in the complex $\omega$ plane is at $\omega=-k$: any other possible poles coming from singularites in $G_{\pm}$ are suppressed by the presence of the delta function, which appeared  because the operator is essentially one-dimensional.

For non-chiral operators the frequency space analysis is more involved, but the analytic structure is well-known: see, e.g., \cite{Son:2002sd} where it was studied in the context of quasi-normal modes of the BTZ black hole. As an example, for spinless operators with $\Delta=2h=2\bar{h}$ non-integer,\footnote{The following expression is, strictly speaking, not defined for $\Delta \in \mathbb{N}$. Nevertheless, the position of the poles can be obtained by analytic continuation of the non-integer case.} the result is simply the generalization of \eqref{eq:BTZ} \cite{PhysRevB.50.258}:
\begin{equation}
G_R^{\Delta,s=0} (\omega,k) \sim   \frac{1}{T^{2(1-\Delta)}} \frac{\Gamma(1-\Delta)}{\Gamma(\Delta)} \frac{\Gamma\left(\frac{\Delta}{2}-i\frac{\omega+k}{4\pi T}\right)\Gamma\left(\frac{\Delta}{2}-i\frac{\omega-k}{4\pi T}\right)}{\Gamma\left(1-\frac{\Delta}{2}-i\frac{\omega+k}{4\pi T}\right)\Gamma\left(1-\frac{\Delta}{2}-i\frac{\omega+k}{4\pi T}\right)}~.
\label{eq:33}
\end{equation}

The singularities are simple poles at $\omega=\pm k - \frac{4\pi i}{\beta}\left(\frac{\Delta}{2}+n\right)$, with $n=0,1,2,...$ away from the real axis. As a consequence $G_R^{\Delta,s=0}(t,x)$ will exponentially decay in time and relax.

\subsection{The simplest 2D CFT: the free boson}

We can confirm the above general results in the $c=1$ theory of a free boson. As a CFT this is a non-trivial theory due to the fact that one can construct vertex operators. On the other hand the fact that the Lagrangian is that of a free Gaussian theory, allows us to also compute all responses using direct quantization rather than CFT methods and this will allow us to check our results in terms of the no-go theorem and our Operator Thermalization Hypothesis.

The Lagrangian of a free 2D boson 
\begin{equation}
L=\frac{1}{2}\int dx \left[(\partial_t \phi(x,t))^2 - (\partial_x \phi(x,t))^2 \right]~,
\end{equation}
has the corresponding Hamiltonian 
\begin{align}
H=\frac{1}{2} \int dx \left[\pi^2 + (\partial_x \phi)^2\right] \\
\pi(x,t)=\partial_t \phi(x,t)~.
\end{align}
To canonically quantize, we decompose the field $\phi(x,t=0)$ and its conjugate momentum $\pi(x,t=0)$ in raising and lowering operators
\begin{align}
\phi(x)&=\int \frac{dp}{2\pi} \frac{1}{\sqrt{2\omega(p)}}\left(a(p) + a(-p)^\dagger\right) e^{ipx} \\
\pi(x)&=\int \frac{dp}{2\pi} (-i) \sqrt{\frac{\omega(p)}{2}}\left(a(p)-a(-p)^\dagger \right) e^{ipx}~.
\end{align}
with $[a(p),a^\dagger(p')]=2\pi \delta(p-p')$. In terms of these operators, the Hamiltonian becomes
\begin{equation}
H=\int \frac{dp}{2\pi} \omega(p) \left(a(p)^\dagger a(p) + \frac{1}{2}[a(p),a(p)^\dagger]\right)~,
\end{equation}
with $\omega(p)=|p|$. From the standard commutators:
\begin{equation}
[H,a(p)^\dagger]=\omega(p) a(p)^\dagger ~~,~~ [H,a(p)]=-\omega(p) a(p)~.
\end{equation}
we can evolve the ladder operators,
\begin{equation}
a(p) (t)=e^{i H t} a(p) e^{-i H t}= a(p) e^{-i\omega(p) t}  ~,~ a(p)^\dagger(t)=e^{iH t} a(p)^\dagger e^{-i H t}=a(p)^\dagger e^{i \omega(p) t}~,
\end{equation} 
so that, the field at time $t$ is
\begin{align}
\phi(x,t)&=\int \frac{dp}{2\pi} \frac{1}{\sqrt{2\omega(p)}}\left(a(p) e^{ipx-i\omega(p) t} + a(-p)^\dagger e^{ipx+i\omega(p) t}\right)~.
\end{align}

It will be useful to impose an IR regulator for many of our computations. Let us therefore consider the theory on a cylinder of size $L$. The momenta and frequencies are then quantized:
\begin{align}
\omega(p)=\frac{2\pi}{L} |k|~, p=\frac{2 \pi}{L} k~, k \in \mathbb{Z}~, \\
a_k=\frac{1}{\sqrt{L}} a(p) ~~,~~ a^\dagger_k=\frac{1}{\sqrt{L}} a^\dagger(p)~,
\end{align}
where we have scaled the creation and annihilation so that $[a_k,a^\dagger_{k'}]=\delta_{kk'}$.
The Hilbert space naturally splits into a tensor product of Hilbert spaces for each oscillator, labelled by $k$. Energy eigenstates can be written as
\begin{align}
|E\rangle=\otimes_k |n_k\rangle~,~~ E=\sum_k n_k \omega_k=\frac{2\pi}{L} \sum_k |k| n_k~,
\end{align}
and the field can be written as 
a right-moving and a left-moving part:
\begin{align}
\phi(x,t)=\sum_{k=0}^\infty \frac{L}{4\pi k}\left(\alpha_k a_k + \alpha^*_k a^\dagger_k +\bar{\alpha}_k a_{-k}+\bar{\alpha}_k^*  a^\dagger_{-k} \right) \\
\alpha_k=\frac{\sqrt{4\pi k}}{L} e^{-\frac{2\pi i}{L} k (t-x)} ~~,~~\bar{\alpha}_k=\frac{\sqrt{4\pi k}}{L} e^{-\frac{2\pi i}{L} k (t+x)} ~~.~~
\end{align}
One needs to be careful when defining the zero-mode operators and coefficients, but this will not affect our results, therefore we skip over those details.
We can define a momentum operator $P(x,t)=\int \frac{dp}{2\pi} p a^\dagger(p) a(p)$. An energy eigenstate is also an eigenstate of that operator, with eigenvalue
\begin{equation}
P=\frac{2\pi}{L}\sum_k k n_k~.
\end{equation}
Note the absence of absolute value, unlike for the energy eigenvalue.

\subsection{Comparison to 2D CFT}

The field $\phi(x,t)$ itself is not a proper conformal operator, but its left (or right)-moving derivative $\partial \phi(x,t)=i(\partial_t-\partial_x)\phi(x,t)$ is.
It is easy to compute the Wightman statistical two-point function: 
\begin{align}
\langle \partial \phi(x,t) \partial \phi(0,0) \rangle_\beta &=\sum_{k=1}^\infty \alpha_k(x,t) \alpha_k^*(0,0) \langle a_k a^\dagger_k \rangle_\beta+ \alpha_k^*(x,t) \alpha_k(0,0) \langle a_k^\dagger a_k \rangle_\beta \nonumber \\
&=\sum_{k=1}^\infty \langle n_k \rangle_\beta \left(\alpha_k(x,t) \alpha^*_k(0,0) + \alpha^*_k(x,t) \alpha_k(0,0) \right)+ \sum_{k=1}^\infty \alpha_k(x,t)\alpha_k^*(0,0) \nonumber \\
&=\frac{4\pi}{L^2}\sum_{k=1}^\infty k \left[\langle n_k\rangle_\beta \left(e^{-\frac{2\pi i k}{L}(t-x)}+e^{\frac{2\pi i k}{L}(t-x)}\right) + e^{-\frac{2 \pi i k}{L}(t-x)}\right]
\label{eq:WightFiniteBeta}
\end{align}
Since the Hilbert space just factorizes into independent sectors for each mode $k$, the modes follow the usual Bose-Einstein distribution:
\begin{align}
\langle n_k \rangle_\beta=\frac{1}{e^{\frac{2\pi |k| \beta}{L}}-1}=\sum_{m=1}^\infty e^{-\frac{2\pi |k| \beta}{L} m} ~.
\end{align}
Interchanging the sums over $m$ and $k$ and defining $d=t-x$, the first term in \eqref{eq:WightFiniteBeta} gives
\begin{align}
\frac{2\pi}{L^2} \sum_{m=1}^\infty \frac{\cos\left(\frac{2\pi(d-im\beta)}{L}\right)+\cos\left(\frac{2\pi(d+im\beta)}{L}\right)-2}{\left(\cos\left(\frac{2\pi d}{L}\right)-\cosh\left(\frac{2\pi m \beta}{L}\right)\right)^2} 
&=-\frac{2}{\pi}\sum_{m=1}^\infty \frac{(d-m\beta)(d+m\beta)}{(d^2+m^2\beta^2)^2}+O(1/L^2)\nonumber \\
&=\frac{1}{\pi d^2} - \frac{\pi}{\beta^2}\frac{1}{\sinh^2\left(\frac{d\pi}{\beta}\right)} + O(1/L^2)~.
\end{align}
The second term, which is also the zero-temperature two-point function, gives
\begin{align}
\langle 0| \partial \phi(x,t) \partial \phi(0,0)|0\rangle&=\sum_{k=1}^\infty \frac{4\pi k}{L^2} e^{-\frac{2\pi i k}{L} d} \nonumber \\
&=-\frac{\pi}{L^2\sin^2\left(\frac{\pi d}{L}\right)} \nonumber \\
&=-\frac{1}{\pi d^2} +O(1/L^2)~.
\end{align}
Note that this sum does not converge unless $-\frac{2\pi i k}{L} d$ has a negative definite real part. Therefore, we should take $d=t-x-i\epsilon$ with $\epsilon\geq 0$  so that our final answer will be well-defined for every $x,t$.\footnote{We should not have divergences other than possible contact divergences at the origin.}
The full answer is
\begin{equation}
\langle \partial \phi(x,t) \partial \phi(0,0) \rangle_\beta=- \frac{\pi}{\beta^2}\frac{1}{\sinh^2\left(\frac{\pi(t-x-i\epsilon)}{\beta}\right)}~.
\end{equation}
Notice that the Wightman function does have an exponential decay here. 
The Wightman function is not the physical response to a perturbation with this operator, however. For that we need the retarded correlation function:
\begin{equation}
G_R(x,t)=-i\Theta(t)\left(\langle \oo(x,t) \oo(0,0)\rangle_\beta - \langle \oo(0,0) \oo(x,t) \rangle_\beta\right)~.
\end{equation}
It's clear from the last line in \eqref{eq:WightFiniteBeta} that the only possible contribution to this difference will come from the zero-temperature two-point function. This is because the terms proportional to $\langle n_k\rangle_\beta$ are the same in $\langle \partial \phi(x,t) \partial \phi(0,0)\rangle$ and in $\langle \partial \phi(0,0) \partial \phi(x,t)\rangle$. More fundamentally, this is because $[\partial \phi(x,t),\partial \phi(0,0)]$ is a c-number, not an operator. Therefore, its expectation value cannot depend on temperature --- indeed it is well known that in free field theories the retarded Green's function of the fundamental field does not depend on the state it is computed in.  The zero-temperature term we need is
\begin{align}
\langle 0|\partial \phi(0,0) \partial \phi(x,t)|0\rangle=\sum_{k=1}^\infty \frac{4\pi k}{L^2} e^{\frac{2\pi i k}{L}(t-x)}~.
\end{align}
Notice that this time we must give $t-x$ a \emph{positive}-definite imaginary part to have convergence, so that our answer will be
\begin{align}
\langle 0|\partial \phi(0,0) \partial \phi(x,t)|0\rangle&=-\frac{1}{\pi(t-x+i\epsilon)^2}+O(1/L^2)
\end{align}
so that the retarded Green's function is
\begin{align}
G_R(x,t)&=\frac{i}{\pi}\Theta(t)\left(\frac{1}{(t-x-i\epsilon)^2}-\frac{1}{(t-x+i\epsilon)^2}\right) \nonumber \\
&=-2 \Theta(t) \delta'(t-x)~.
\end{align}

The retarded Green's function for $\oo=\partial\phi(x,t)$ thus has no exponential decay and this operator does not thermalize.
This is what we expected on general CFT grounds. The operator $\partial\phi$ is a chiral operator with $h=1,\bar{h}=0$ and chiral operators do not thermalize. The advantage of also understanding this result from direct quantization, is that we can now try to understand this in terms of our no-go condition. Consider the matrix element
\begin{equation}
\langle \otimes_k n'_k|\partial \phi  | \otimes_k n_k\rangle= \sum_{l=1}^\infty \langle \otimes_k n'_k|  \alpha_l a_l + \alpha^*_l a^\dagger_l |\otimes_k n_k\rangle~.
\end{equation}
It is clear the summand on the right-hand-side is only non-zero if all the occupation numbers match except for $k=l$, whose occupation number must differ by exactly one. The difference in momentum between these states $|\otimes_k n_k\rangle$  and $|\otimes_{k\neq l} n_k;n_l=n_k\pm 1\rangle$ is $\frac{2\pi}{L}\sum_{k} k (n_k-(n_k\pm \delta_{k,l}))=\pm \frac{2\pi}{L}{l}$, and the difference in energy is $\frac{2\pi}{L}\sum_{k} |k| (n_k-(n_k\pm \delta_{k,l}))=\pm \frac{2\pi}{L} l$. Since the difference in energies is fixed by the difference in momenta, there can be no dissipation according to our no-go theorem.

The confirmation of our no-go result for chiral operators $\partial \phi$ is in fact readily extended to any chiral primary operator in a two-dimensional CFT. To be precise, consider the Euclidean time CFT on a cylinder with circumference $L$. The spectrum is discrete at finite $L$, but we shall take a thermodynamic limit in the end. By the usual radial quantization procedure, we can do a conformal transformation from the cylinder to the plane. The Hamiltonian on the cylinder is mapped to $\frac{2\pi}{L}(L_0+\bar{L}_0)$ and the momentum operator to $\frac{2\pi}{L}(L_0 - \bar{L}_0)$. When restricted to the (anti)-chiral i.e. (anti-)holomorphic sector, we therefore have $H=(-)P$. Given two states $|h_1,\bar{h}_1\rangle$ and $|h_2,\bar{h}_2\rangle$, the matrix elements of a chiral operator will always be of the form $\langle h_1,\bar{h}_1|\oo_{chiral}|h_2,\bar{h}_2\rangle=\delta_{\bar{h}_1,\bar{h}_2} \langle h_1|\oo_{chiral}|h_2\rangle$. But given that $\Delta E =\frac{2\pi}{L} (\Delta h+\Delta \bar{h})$ and $\Delta P =\frac{2\pi}{L} (\Delta h - \Delta \bar{h}) $ the nonzero value of $\langle h_1,\bar{h}_1|\oo_{chiral}|h_2,\bar{h}_2\rangle$ will always have $\Delta P=\Delta H$ and our no-go condition will always be satisfied.
The thermodynamic limit $L\rightarrow \infty$, where we find previously quoted universal finite-temperature two-point functions, cannot induce any violations of this property and hence establishes the no-go condition for chiral operators.

To check the non-chiral case, consider a composite operator \mbox{$C(x,t)=\oo_1(x+\zeta,t) \oo_2(x,t)$}, with $\zeta\geq 0$ infinitesimal picked to avoid any contact divergences. The retarded Green's function for such an operator is
\begin{align}
G_R^C(x,t)&= -i\Theta(t)\langle [C(x,t),C(0,0)]\rangle_\beta 
\label{eq:compop} \\
&=-i\Theta(t) \left[\langle \oo_1(x+\zeta,t) \oo_1(\zeta,0) [\oo_2(x,t),\oo_2(0,0)] + \oo_1(x+\zeta,t)[\oo_2(x,t),\oo_1(\zeta,0)]\oo_2(0,0)\rangle_\beta \nonumber \right.\\
&\left. +\langle \oo_1(\zeta,0)[\oo_1(x+\zeta,t),\oo_2(0,0)]\oo_2(x,t)+[\oo_1(x+\zeta,t),\oo_1(\zeta,0)]\oo_2(0,0)\oo_2(x,t)\rangle_\beta \right]~. \nonumber
\end{align}
If we pick $\oo_1=\partial \phi$ and $\oo_2=\bar{\partial} \phi =i(\partial_t+\partial_x)\phi(x,t)$, where for the latter we have 
\begin{align}
\langle \bar{\partial}\phi(x,t) \bar{\partial}\phi(0,0)\rangle_\beta &=-\frac{\pi}{\beta^2}\frac{1}{\sinh^2\left(\frac{\pi(t+x-i\epsilon)}{\beta}\right)} \\ 
G_R(x,t)&=-\frac{i}{\pi} \Theta(t) \left(\frac{1}{(t+x-i\epsilon)^2}-\frac{1}{(t+x+i\epsilon)^2} \right)~.
\end{align} 
then the retarded Green's function becomes:
\begin{align}
G_R^{:\partial\phi\bar{\partial}\phi:}(x,t)=\Theta(t)\frac{2}{\beta^2}\left[\frac{1}{\sinh^2\left(\frac{\pi}{\beta}(t-x-i\epsilon)\right)}\delta'(t+x)+\frac{1}{\sinh^2\left(\frac{\pi}{\beta}(t+x+i\epsilon)\right)}\delta'(t-x)\right]
\end{align}
Again away from $|x|=|t|$ the response is zero, but on either $x=-t$ or $x=t$ the function decays exponentially in time.

It will be illustrative to compare this to the chiral composite operator $C = :\partial \phi\partial \phi:$ for which
\begin{equation}
G_R^{:\partial\phi{\partial}\phi:}(x,t)=\Theta(t)\frac{4}{\beta^2}\delta'(t-x)\left[\frac{1}{\sinh^2\left(\frac{\pi}{\beta}(t-x-i\epsilon)\right)}+\frac{1}{\sinh^2\left(\frac{\pi}{\beta}(t-x+i\epsilon)\right)}\right]
\end{equation}
As we by now know it should be, the chiral operator $C=:\partial \phi \partial \phi:$ does not thermalize as it obeys the no-go theorem.

We can now also check the converse whether the non-chiral operator satisfies the Operator Thermalization Hypothesis using direct quantization.

Comparing the two composite operators, we have
\begin{align}
:\bar{\partial}\phi \partial \phi:&=\sum_{k,k'=1}^\infty \left[ \alpha_k \bar{\alpha}_{k'} a_k a_{-k'} + \alpha_k \bar{\alpha}^*_{k'} a_k a^\dagger_{-k'} + \alpha_k^* \bar{\alpha}_{k'} a^\dagger_k a_{-k'} + \alpha^*_k \bar{\alpha}^*_{k'} a^\dagger_k a^\dagger_{-k'}\right] \\
:\partial \phi \partial \phi:&=\sum_{k,k'=1}^\infty \left[ \alpha_k \alpha_{k'} a_k a_{k'} + 2\alpha_k \alpha^*_{k'}a^\dagger_{k'} a_k  + \alpha^*_k \alpha^*_{k'} a^\dagger_k a^\dagger_{k'}\right]~.
\end{align}
Consider each of the three different kinds of terms individually. First, take the term with two annihilation operators. For the non-chiral operator, we have:
\begin{equation}
\langle \otimes_l n_l|a_k a_{-k'}| \otimes_l n'_l\rangle=\left(\prod_{l\neq {k,k'}} \delta_{n_l n'_l}\right) \left(\delta_{n_k,n'_k+1}\right)\left( \delta_{n_{-k'}, n'_{-k'}+1}\right)\sqrt{n'_k n'_{-k}}~.
\end{equation}
This is non-zero if and only if the momentum difference between the bra and the ket is $P'-P=\frac{2\pi}{L}(k-k')$. The energy difference between these two states, however, is $E'-E=\frac{2\pi}{L}(k+k')$. We can see that the energy and momentum difference can be adjusted independently while still giving a non-zero matrix element.
By contrast, if we consider $a_k a_{k'}$ (the analogous term in the chiral operator), we get:
\begin{align}
\langle \otimes_l n_l|a_k a_{k'}| \otimes_l n'_l\rangle&=(1-\delta_{kk'})\left(\prod_{l\neq {k,k'}} \delta_{n_l n'_l}\right) \left(\delta_{n_k,n'_k-1}\right)\left( \delta_{n_{k'} n'_{k'}-1}\right) \sqrt{n'_k n'_{k'}} \nonumber \\ 
&+  \delta_{kk'}\left(\prod_{l\neq {k}} \delta_{n_l n'_l}\right) \left(\delta_{n_k,n'_k-2}\right) \sqrt{n'_k (n'_k-1)} ~.
\end{align}
This time, the momentum difference between the states must be $P'-P=\frac{2\pi}{L}(k+k')$, but now this also equals the energy difference. 
Similar arguments apply for the $a^\dagger a$ and $a^\dagger a^\dagger$ terms, but this is already sufficient to state that the non-chiral operator violates our no-go theorem. We also know that the non-chiral operator thermalizes whereas the chiral one does not, and this therefore verifies our OTH.

The above example illustrates a deeper principle behind our no-go theorem and the OTH. A very naive guess for operator thermalization in integrable theories could have been that operators that carry only a few of the infinite set of conserved charges do not thermalize --- e.g. exciting a single momentum mode --- but ones that change a macroscopically large number do. 
The thermalization of the operator $\oo = :\partial\phi\bar{\partial}\phi:$ clearly shows that this intuition is incorrect.

\subsection{Vertex operators}
In addition to $\partial \phi, \bar{\partial}\phi$ and (normal ordered) products thereof, there is another natural kind of conformal operator we can build out of creation and annihilation operators in the $c=1$ free boson theory: vertex operators. These are defined as
\begin{equation}
V_{\xi}(x,t)=:e^{i \xi \phi(x,t)}:~.
\end{equation}
Once again, in direct quantization normal-ordering is to put the annihilation operators to the right. These operators are not Hermitian, but it is of course a simple matter to construct two linear combinations that are, and can be used to perturb the Hamiltonian:
\begin{equation}
:\cos(\xi \phi):\equiv \frac{V_\xi+V^\dagger_{\xi}}{2}~,~:\sin(\xi \phi):\equiv \frac{V_\xi-V^\dagger_{\xi}}{2i}
\end{equation}

It is instructive to see how such operators, which interact with conserved charges in a much more complicated way, thermalize. Let us also define
\begin{align}
V^{(\pm)}_{\xi}(x,t)=:e^{i \xi \phi^{(\pm)}(x,t)}:~~~~,~
V_\xi(x,t)=V^{(-)}_{\xi}(x,t) V^{(0)}_{\xi} V^{(+)}_\xi(x,t)~,
\end{align}
where $\phi^{(\pm,0)}(x,t)$ contains only the positive, negative or zero momentum modes. In other words, $V^{\pm}_{\xi}$ is the (anti)holomorphic part of $V_\xi$ (the zero mode will contribute only an overall constant to correlation functions). Based on our arguments of the previous section, operators built out of linear combinations of $V^{(\pm)}_\xi$ cannot thermalize, even though they appear to be spread over a much larger fraction of the Hilbert space than $\bar{\partial} \phi \partial \phi$ and they change not a microscopic but a macroscopic  number of  the conserved charges in the system. Nevertheless, we will see that we can understand the thermalization of $V_{\xi}$ but the absence of thermalization of $V^{(\pm)}_{\xi}$ in terms of our no-go theorem.

To do so we compute finite-temperature correlators of the vertex operators $V_{\xi}$ and  $V_{\xi}^{(\pm)}$ and their Hermitian conjugates.
For $A_1$,$A_2$ linear combinations of creation and annihilation operators, it is easy to show the operator identity 
\begin{equation}
\langle\psi| :e^{A_1}: :e^{A_2}:|\psi \rangle= \langle \psi |:e^{A_1+A_2}:|\psi \rangle e^{\langle 0|A_1 A_2 |0\rangle}~.
\end{equation}
Then the Wightman function of the vertex operator is obtained by picking $A_1=-i\xi\phi(x,t)$ and $A_2=i\xi\phi(0,0)$.
\begin{align}
\langle :V_\xi^\dagger(x,t): :V_\xi(0,0): \rangle_\beta&=\langle :e^{A_1+A_2}:\rangle_\beta e^{\langle 0| A_1 A_2|0\rangle}\nonumber \\
A_1+A_2&=\sum_{k=1}^\infty (\gamma_k a_k - \gamma_k^* a^\dagger_k + \bar{\gamma}_k a_{-k}-\bar{\gamma}_k^* a^\dagger_{-k}) \nonumber \\
\gamma_k&=\frac{i\xi}{\sqrt{4\pi k}}\left[1-e^{-\frac{2\pi i}{L} k (t-x)}\right]~,~ \bar{\gamma}_k=\frac{i\xi}{\sqrt{4\pi k}}\left[1-e^{-\frac{2\pi i}{L} k (t+x)}\right]~.
\end{align}
We first compute $\langle 0|A_1 A_2|0\rangle$. The computation is textbook: the non-vanishing terms are:
\begin{align}
\langle 0|A_1 A_2|0\rangle&=C+ \xi^2 \sum_{k=1}^\infty \frac{e^{-\frac{2\pi i}{L} k (t-x)}+e^{-\frac{2\pi i}{L} k (t+x)}}{4\pi k} \nonumber \\
&=C-\frac{\xi^2}{4\pi}\left[ \log\left(1-e^{-\frac{2\pi i}{L}(t-x-i\epsilon)}\right)+\log\left(1-e^{-\frac{2\pi i}{L}(t+x-i\epsilon)}\right)\right]  
\end{align}
where $C$ is a real constant accounting for the zero-mode contribution and we have again inserted an $i\epsilon$ to ensure the convergence of the sums. Exponentiating, we have
\begin{align}
e^{\langle 0| A_1 A_2|0 \rangle} &= C' \left[2ie^{-i\frac{\pi}{L}(t-x-i\epsilon)}\sin\left(\frac{\pi(t-x-i\epsilon)}{L}\right)\right]^{\frac{-\xi^2}{4\pi}} \left[2i e^{-i\frac{\pi}{L}(t+x-i\epsilon)}\sin\left(\frac{\pi(t+x-i\epsilon)}{L}\right)\right]^{\frac{-\xi^2}{4\pi}} \nonumber\\
&\sim\frac{1}{\left[\frac{2i\pi (t-x-i\epsilon)}{L}\right]^\frac{\xi^2}{4\pi}\left[\frac{2i\pi (t+x-i\epsilon)}{L}\right]^\frac{\xi^2}{4\pi}}~,
\end{align}
Note that terms of the form $\langle V_\xi(x,t)V_\xi(0,0) \rangle$ are subleading in $1/L$. We therefore ignore them when calculating the Green's function for $:\cos(\xi \phi):$ and $:\sin(\xi \phi):$.
For the finite temperature response we also need $\langle :e^{A_1+A_2}: \rangle_{\beta}$.
The details of the calculation can be found in the Appendix. We quote here the result:
\begin{equation}
\langle :e^{A_1+A_2}: \rangle_\beta= K \left(\frac{\pi(t+x-i\epsilon)}{\beta\sinh\left(\frac{\pi (t+x-i\epsilon)}{\beta}\right)}\right)^{\frac{\xi^2}{4\pi}} \left(\frac{\pi(t-x-i\epsilon)}{\beta\sinh\left(\frac{\pi (t-x-i\epsilon)}{\beta}\right)}\right)^{\frac{\xi^2}{4\pi}}~,
\end{equation}
where $K$ is an overall real constant. With this, we can write down the  finite- temperature expectation value for the Wightman function of the vertex operator:
\begin{equation}
\langle V_\xi^\dagger(x,t) V_\xi(0,0)\rangle_\beta =  \left(\frac{\pi^2 L^2}{4\beta^2 \sinh\left(\frac{\pi(t-x-i\epsilon)}{\beta}\right)\sinh\left(\frac{\pi(t+x-i\epsilon)}{\beta}\right)}  \right)^{\frac{\xi^2}{4\pi}} ~. 
\end{equation}
With these ingredients, we can read off the holomorphic and anti-holomorphic scaling dimensions of both $:\cos(\xi \phi):$ and $:\sin(\xi \phi):$ to be:
\begin{equation}
(h,\bar{h})=(\frac{\xi^2}{8\pi},\frac{\xi^2}{8\pi})~. \label{eq:dim-v}
\end{equation} 
Using the same general analysis as before, we see that these operators thermalize, as expected.\footnote{The case of $\xi^2=4\pi n$ with $n\in \mathbb{N}$ requires a more careful treatment. It corresponds to the operators being spinless primaries with integer scaling dimension $\Delta=n$, which we have seen previously is more subtle but still exhibits exponential decay at long times.}

In contrast, consider the case of $V^{(+)}_{\xi}(x,t)$. By picking $A_1^{(+)}=-i\xi \phi^{(+)}(x,t)$ and $A_2^{(+)}=i\xi \phi^{(+)}(0,0)$, we get that
\begin{equation}
\langle V_\xi^{(+)\dagger}(x,t) V_\xi^{(+)}(0,0)\rangle_\beta=\langle :e^{A_1^{(+)}+A_2^{(+)}}:\rangle_\beta e^{\langle 0|A_1^{(+)} A_2^{(+)}|0\rangle}~.
\end{equation}
However, by construction $\phi^{(+)}(x,t)$ depends only on $t-x$, and therefore the same cancellation in the retarded Green's function will happen when $t\neq x$, preventing thermalization.

We can now check this against our no-go theorem by looking at the matrix elements of the operators in question in terms of direct quantization.
Consider two energy eigenstates, labelled by the occupation number of each mode $k$: $|\otimes_k n_k\rangle$ and $|\otimes_k n'_k\rangle$. We can write
\begin{align}
:e^{i\xi\phi}:&=\prod_k e^{i \xi \mu_k^* a^\dagger_k} e^{i \xi \mu_k a_k}\\
&=\prod_k e^{-\xi^2 |\mu_k|^2} e^{i\xi\mu_k a_k} e^{i \xi \mu_k^* a^\dagger_k}~,
\end{align}
therefore we have
\begin{align}
\langle \otimes_k n'_k|:e^{i \xi \phi}:|\otimes_k n_k\rangle &= \prod_k e^{-\xi^2 |\mu_k|^2} \langle n'_k|e^{i \xi \mu_k a_k} e^{i \xi \mu_k^* a^\dagger_k} |n_k\rangle \\
&=\prod_k e^{-\xi^2 |\mu_k|^2} \langle 0_k|e^{i\xi \mu_k a_k} \frac{(a_k)^{n'_k}}{\sqrt{n'_k!}} \frac{(a^\dagger_k)^{n_k}}{\sqrt{n_k!}}e^{i \xi \mu_k^* a^\dagger_k}|0_k\rangle~.
\end{align}
Recall that coherent states of a harmonic oscillator take the form:
\begin{align}
e^{\alpha a^\dagger}|0\rangle=e^{\frac{|\alpha|^2}{2}}|\alpha\rangle=\sum_{l=0}^\infty \frac{\alpha^l}{\sqrt{l!}}|l\rangle~,
\end{align}
so that
\begin{align}
e^{-\xi^2 |\mu_k|^2}\langle n'_k|e^{i \xi \mu_k a_k} e^{i \xi \mu_k^* a^\dagger_k} |n_k\rangle &= \sum_{l_k,l'_k} \langle l'_k|\frac{(i\xi \mu_k)^{l_k'}}{\sqrt{l_k'!}} \frac{(a_k)^{n'_k}}{\sqrt{n'_k!}} \frac{(a^\dagger_k)^{n_k}}{\sqrt{n_k!}}\frac{(i\xi\mu_k^*)^{l_k}}{\sqrt{l_k!}}|l_k\rangle   \nonumber \\
&= \sum_{l_k ,l'_k} \frac{(i\xi\mu_k)^{l_k'}}{\sqrt{l_k'!}} \frac{(i\xi\mu_k^*)^{l_k}}{\sqrt{l_k!}} \frac{1}{\sqrt{n_k! n'_k!}} \sqrt{\frac{(l'_k+n'_k)!(l_k+n_k)!}{l'_k! l_k!}} \delta_{l_k+n_k,l'_k+n'_k} \nonumber \\
&=\begin{cases}
      (i\xi \mu_k)^{n_k-n'_k} \sqrt{\frac{n_k!}{n'_k!}} {}_1F^{(R)}_1\left(1+n_k,1+n_k-n'_k;-\xi^2|\mu_k|^2\right), & \text{if}\ n_k \geq n'_k \\
       (i\xi \mu_k^*)^{n_k'-n_k} \sqrt{\frac{n'_k!}{n_k!}} {}_1F^{(R)}_1\left(1+n'_k,1+n'_k-n_k;-\xi^2|\mu_k|^2\right), & \text{otherwise}
    \end{cases}~ \nonumber \\
& \equiv c(n_k,n'_k ; \xi)~.    
\end{align}
$_1F^{(R)}_1(a,b;z)\equiv {}_1F_1(a,b;z)/\Gamma(b)$ is a regularized hypergeometric function. This leads to
\begin{align}
\langle \otimes n'_k|V_\xi|\otimes n_k\rangle&=\prod_k c(n_k,n'_k ; \xi)\\
\langle \otimes n'_k|:\cos(\xi \phi):|\otimes n_k\rangle&=\frac{1}{2}\left(\prod_k c(n_k,n'_k;\xi)+\prod_k c(n_k,n'_k;-\xi)\right) \\
\langle \otimes n'_k|:\sin(\xi \phi):|\otimes n_k\rangle&=\frac{1}{2i}\left(\prod_k c(n_k,n'_k;\xi)-\prod_k c(n_k,n'_k;-\xi)\right)~.
\end{align}

Note that none of these factors are zero, therefore none of the matrix elements of $V_{\xi} = :e^{i\xi\phi}:$ are zero. 

Notice that $c(n_k,n'_k;\xi)=(-1)^{(n_k-n'_k)} c(n_k,n'_k;\xi)$, so that the matrix elements of $:\cos(\xi \phi):$ only vanish for states where $\sum_k n_k-n'_k$ is odd (and those of $:\sin(\xi \phi):$ when that sum is even). This is not sufficient to ensure that the only non-zero matrix elements have $\Delta E$ given by a function of $\Delta P$, therefore both of these operators fail our no-go condition.

Conversely, it is straightforward to see that for a chiral vertex operator, say $V_{\xi}^{(+)}$, the matrix elements will be zero whenever $n'_{-\tilde{k}}\neq n_{-\tilde{k}}$ for at least one $\tilde{k}>0$. This is because that operator contains only positive-momentum modes. As a consequence the energy difference between the states is equal the momentum difference for all non-zero matrix elements, satisfying our no-go condition, and the operator does not thermalize.

\section{The Transverse Field Ising model}

Having seen that it is the nature of the operator that determines whether it relaxes away into the thermal state or not, we implicitly know the answer to the paradox that we raised in the introduction. Famously the magnetization in the transverse field Ising model at the critical point thermalizes. However, the Ising model at the critical point is also equivalent to a free fermion theory. It should be that the magnetization indeed fails our no-go theorem and should relax according to our Operator Thermalization Hypothesis. We will now show that this is so.

The transverse field Ising model has the following Hamiltonian (see, e.g., \cite{sachdev_2011}):
\begin{equation}
H=-J\sum_{i=1}^N \left(g\sigma_i^x + \sigma_i^z \sigma_{i+1}^z\right)~,
\end{equation}
with $[\sigma_i^a,\sigma_j^b]=2i\epsilon_{abc} \delta_{ij}$.
In the thermodynamic limit, $g=1$ is a critical point. At that point (at zero temperature), the system undergoes a second order quantum phase transition from an ordered ($g>1$) to disordered ($g<1$) phase. As such, the theory at $g=1$ is well described by a CFT. This means we can immediately apply our CFT insights from the previous section.  

Moreover, the CFT is a free fermion theory. This equivalence follows after a Jordan-Wigner transformation of the spins:
\begin{equation}
\s_i^x=1-2c_i^\dagger c_i ~~,~~ \s_i^z=-\prod_{j<i}(1-2c_j^\dagger c_j)(c_i+c_i^\dagger)~, \label{eq:JW}
\end{equation}
The fermionic modes $c_i$ obey the conventional anti-commutation relations:
\begin{align}
\{c_i,c_j^\dagger \}&=\delta_{ij}~,\\
\{c_i,c_j\}&=\{c_i^\dagger,c_j^\dagger\}=0~.
\end{align} 
Going to momentum space, $c_k=\frac{1}{\sqrt{N}}\sum_{j} c_j e^{-ikr_j}$, the JW-transformed Hamiltonian can be rewritten as 
\begin{equation}
H=J\sum_k \left[2(g-\cos(ka))c_k^\dagger c_k + i \sin(ka)(c_{-k}^\dagger c_{k}^\dagger + c_{-k} c_k)-g\right]~,
\end{equation}
where $a$ is the lattice spacing.
Finally, a Bogoliubov transform
\begin{align}
c_k&=\cos(\theta_k/2)\gamma_k + i \sin(\theta_k/2)\gamma_{-k}^\dagger~,
\end{align}
with $\theta_k$ defined by
\begin{align}
\tan \theta_k=\frac{\sin(ka)}{g-\cos(ka)}~,
\end{align}
 diagonalizes the Hamiltonian to that of a free fermion theory
\begin{align}
H&=\sum_k \epsilon_k \left(\g^\dagger_k \g_k -\frac{1}{2}\right)~.
\end{align}
with dispersion relation
\begin{align}
\epsilon_k\equiv2J(g\cos\theta_k - \cos(\theta_k+ka))&=2J\sqrt{1+g^2-2g\cos(ka)}~.
\end{align}
As the reformulated theory reveals that the theory is free, the spectrum is straightforward. Eigenstates of the Hamiltonian are simply labelled by the occupation number (0 or 1) of each momentum mode. In other words, the energy eigenstates are also eigenstates of $\gamma^\dagger_k \gamma_k$ for each value of $k$. By mapping the theory to that of a free fermion, we see explicitly that it is integrable.

At zero momentum, we have 
\begin{equation}
\epsilon_{k=0}=2J|g-1|~,
\end{equation}
so that the theory becomes gapless as $g\rightarrow 1$. For $g=1$, the exact dispersion relation is
\begin{equation}
\epsilon_k=4J\left|\sin\left(\frac{ka}{2}\right)\right| = 2 J a |k| +\mathcal{O}(ka)^2~.
\end{equation} 
For momentum small compared to the inverse lattice constant, this is the dispersion relation of massless relativistic particles with the ``speed of light'' being $2Ja$.

In the continuum limit, we can do a gradient expansion. Defining \cite{sachdev_2011}
\begin{equation}
\Psi(x_i)=\frac{1}{\sqrt{a}} c_i~,
\end{equation}
the Hamiltonian becomes
\begin{equation}
H=E_0+\int dx\left[\frac{v}{2}\left(\Psi^\dagger \frac{\partial \Psi^\dagger}{\partial x} - \Psi \frac{\partial \Psi}{\partial x}\right)+ m \Psi^\dagger \Psi\right]+\cdots~,
\end{equation}
with
\begin{equation}
m=2J(1-g)~,~v=2Ja~,
\end{equation}
where $E_0$ is a constant. 
If one defines a pair of Majorana fermions $\psi_{1}$ and $\psi_{2}$ as:
\begin{align}\label{com}
    \psi_{1}=\frac{\Psi+\Psi^{+}}{\sqrt{2}}, \ \ \ \ \ \psi_{2}=i\frac{\Psi-\Psi^{+}}{ \sqrt{2}}, \ \ \ \ \  \{ \psi_{1}, \psi_{2} \} = 0, \ \ \ \ \psi_{1}^{2}=\psi_{2}^{2}=\frac{1}{2a},
\end{align}
and then performs a rotation in the space of $\psi_{1}$ and $\psi_{2}$:
\begin{align}\label{majoranas_def}
    \psi_{+}=\frac{\psi_{1}+\psi_{2}}{\sqrt{2}}, \ \ \ \ \ \ \psi_{-}=\frac{\psi_{2}-\psi_{1}}{\sqrt{2}},
\end{align}
the Euclidean action in terms $\psi_{+}$ and $\psi_{-}$  is written as follows:
\begin{align}
    S_{E}=\frac{1}{2}\int dx d \tau \ \left( \psi_{+} \left( \partial_{\tau}+ i v \partial_{x} \right) \psi_{+} + \psi_{-} \left( \partial_{\tau} - i v\partial_{x}\right)\psi_{-} +  2i m \psi_{-} \psi_{+} \right).
\end{align}
At the critical point, the mass is 0  and we're left with a two-dimensional conformal field theory of two (non-interacting) Majorana fermions. Here, $\psi_{-}$ and $\psi_{+}$ are holomorphic and antiholomorphic, respectively.

We immediately deduce that the (anti-)holomorphic operators $\psi_{-}$ and $\psi_{+}$ do not thermalize. As all chiral operators, they obey the no-go condition. However, it is clear from the form of the Jordan-Wigner transformation \eqref{eq:JW} that $\sigma^{z}_i$ will be mapped to a rather non-trivial composite operator.
\begin{align}
\sigma_z(x) \sim e^{ \pi \int_{-\infty}^x\psi_{-}(y)\psi_{+}(y) dy}\left(\psi_{+}(x)-\psi_{-}(x)\right)
\end{align}
An important point is that this operator is non-local. Non-local operators are traditionally excluded from much of the usual studies of thermalization as they encode instantaneous correlations on wide scales that can hide local dynamics leading to relaxation. However, at the critical point in the presence of conformal symmetry this cannot be used as an argument. Many naively non-local operators are equivalent to  local operators after a conformal transformation.   

We are precisely interested in the theory at the critical point where the perturbation is local. This was the objective of the original calculation by Damle and Sachdev \cite{Damle:1997rxu}. At this point, $\sigma_z$ will correspond to a scalar operator with $h=\bar{h}=\frac{1}{16}$: this is the famous anomalous dimension of the order parameter of the Ising model. It can be established by computing the auto-correlation function of $\sigma^z_i$ directly from the Jordan-Wigner transformation in the scaling limit \cite{PhysRevA.3.786,PERK19841, sachdev_2011} or by using the spin-disorder duality of the model \cite{Belavin:1984vu}. 
An illustrative approach is to use bosonization: all correlation functions of $\sigma^z_i$ can be obtained from the square root of the correlation functions of a vertex operator in a free scalar field theory 

The transverse field Ising model cannot be bosonized straightforwardly since its central charge is $1/2$ and the central charge of a free bosonic theory is 1. We will follow the bosonization procedure described in \cite{Zuber:1976aa, gogolin2004bosonization}. The main trick is to consider two non-interacting copies of the Ising model so that the total action is the action of a massless Dirac fermion:
\begin{align}
    S= 2 \int d\tau d x \left( R^{+} \partial_{\bar{z}} R + L^{+}\partial_{z} L \right),
\end{align}
where $z=\tau+\frac{i x}{v}$, $\partial_{z} = \frac{1}{2} \left(\partial_{\tau}- i v \partial_{x} \right)$. From now on, we will set $v=1$ for convenience.
$R$ and $L$ can be represented in terms of the Majorana fermions as:
\begin{align}\label{rl}
    R= \frac{\psi_{+}^{1}+i\psi_{+}^{2}}{\sqrt{2}}, \ \ \  L = \frac{\psi_{-}^{1}+i\psi_{-}^{2}}{\sqrt{2}},
\end{align}
where the indices 1 and 2 denote the first and the second copy of the Ising model. R and L can be bosonized as follows:
\begin{align}\label{rules_bs}
    R &\rightarrow \frac{1}{\sqrt{2\pi a}} \exp{\left(i \sqrt{4\pi}\bar{\phi}(\bar z)\right)} \nonumber \\
    L &\rightarrow \frac{1}{\sqrt{2\pi a}} \exp{\left(- i \sqrt{4\pi}\phi( z)\right)} .
\end{align}
 $\phi(z)$ and $\bar{\phi}(\bar {z})$ are the holomorphic and antiholomorphic parts of the bosonic field $\Phi(z,\bar{z})=\bar{\phi}(\bar {z})+ \phi (z)$ with the action:
\begin{align}
    S=\frac{1}{2} \int d \tau d x \left( (\partial_{\tau} \Phi)^{2}+(\partial_{x} \Phi)^{2} \right).
\end{align}
The currents of the right- and left-moving fermions are:
\begin{align}\label{currr}
R^{+}R = -\frac{i}{\sqrt{\pi}}\partial_{\bar{z}} \bar{\phi}(\bar{z}) \\
L^{+}L = \frac{i}{\sqrt{\pi}}\partial_{z}\phi(z).
\end{align}
Now, consider the square of the spin-spin correlator of a single copy of the Ising chain:
\begin{align}\label{square}
       \left( \langle \sigma^{1}_{z} (x) \sigma^{1}_{z}(0) \rangle\right)^{2}=\langle \sigma^{1}_{z}(x) \sigma^{1}_{z}(0) \sigma^{2}_{z}(x) \sigma^{2}_{z} (0) \rangle \stackrel{\mathrm{def}}{=}  \langle \sigma_{z} (x) \sigma_{z}(0)\rangle.
\end{align}

Using the bosonization rules, one can show explicitly that, in the continuum limit,
\begin{align}
      \langle \sigma_{z} (x) \sigma_{z}(0)\rangle  =\frac{1}{\pi^{2}}\left \langle :\sin{\sqrt{\pi} \Phi(x)}: :\sin{\sqrt{\pi}\Phi(0)}: \right\rangle.
\end{align}
As a result, one can represent $\sigma_{z}(x) = \sigma^{1}_{z}(x) \sigma^{2}_{z}(x)$ in the doubled Ising model as
\begin{align}
    \sigma_{z}\sim :\sin{\sqrt{\pi}\Phi(x)}:
\end{align}
and compute the correlation function $\langle \sigma_{z}(x) \sigma_{z}(0) \rangle$. To obtain its value for the physical model, one has to take the square root. It can be seen from \eqref{eq:dim-v} that the conformal dimensions of $\sigma_{z} = \sigma^{1}_{z}\sigma^{2}_{z}$ are $(h,\bar{h})=(1/8,1/8)$. This of course lets us recover the known result that the magnetization of the transverse field Ising thermalizes. Moreover, that result is recovered precisely through a free field theory calculation.

With this equivalence, we can immediately apply our results from the previous section to explain why $\sigma^1_z$ thermalizes even though the underlying theory is free.

\section{Higher-dimensional field theory}

To show the role of conserved charges and the fact that our previous results are not purely the consequence of the constrained kinematics of $1+1$-dimensional theories, we briefly discuss the case of a charged scalar field in $d$ dimensions. Such a field has the following Lagrangian:
\begin{equation}
L=\frac{1}{2}\int d^{d-1}x \left(\partial_t \phi \partial_t \phi^* - \nabla \phi \nabla \phi^* - m^2 \phi \phi^*\right)~.
\end{equation}
Expanding the field $\phi$ in creation and annihilation operators
\begin{align}
\phi(x,t)=\int \frac{d^{d-1}p}{(2\pi)^{d-1}} \frac{1}{\sqrt{2\omega(p)}} \left(a_{\vec{p}}  e^{-i p \cdot x} + \bar{a}^\dagger_{\vec{p}} e^{i p \cdot x}\right)~,
\end{align}
the energy eigenstates are then labelled by the occupation number of both sets of modes. As discussed in the Introduction, the scalar field itself does not thermalize, even at finite temperature. This is easy to see from our no-go condition: if $\langle E_i|a_{\vec{p}}|E_j\rangle \neq 0$, then we must have $\vec{P}_i-\vec{P}_j=\vec{p}$ and $E_i-E_j=\sqrt{\vec{p} \cdot\vec{p}+m^2}$ (and similarly for the daggered and barred operators).
Note that, as in the lower-dimensional case, the commutators $[\phi(x),\phi(y)]$, $[\phi(x),\phi^*(y)]$ and their complex conjugates are all c-numbers. The retarded Green's function of $\phi(x)$\footnote{Or of $\text{Re}[\phi]$, $\text{Im}[\phi]$, if we are interested in Hermitian operators.} therefore does not depend on the state. It is given by
\begin{align}
G_R(x)=-i \Theta(t) \langle [\phi^*(x),\phi(0)]\rangle_\beta=i \Theta(t) \int \frac{d^{d-1}p}{(2\pi)^{d-1}} \frac{1}{2\omega(p)} \left(e^{ip\cdot x}-e^{-ip \cdot x}\right)
\label{eq:gr-op}
\end{align}
and clearly does not thermalize. 
Note that this independence of the state only holds for the retarded Green's function. The finite-temperature Feynman propagator, that one needs to compute perturbative corrections in an interacting theory, does depend on temperature. For a symmetric Schwinger-Keldysh contour, it equals
\begin{align}
G_F(\vec{k},\omega)=\text{Re}\left[G_R(\vec{k},\omega)\right]+i\coth\left(\frac{\beta \omega}{2}\right)\text{Im}\left[G_R(\vec{k},\omega)\right] 
\end{align}
Compared to the retarded Green's function the Feynman Green's function has an additional series of poles at $\omega=\frac{2\pi i n}{\beta}$, $n\in\mathbb{Z}$.
The Feynman propagator also comes into play when computing responses of composite operators. It is the exponential relaxation associated with these poles that will lead to the thermalization of such composite operators. Consider for example, the ``mass'' operator
\begin{equation}
\oo(x,t)\equiv \phi^\dagger(x,t) \phi(x,t)~.
\end{equation}

Using the decomposition \eqref{eq:compop} for the composite operator and using that for $t>0$ $\langle \phi^\dagger(x,t) \phi(0,0)\rangle_\beta=-iG_F(x,t)$, we can write:
\begin{equation}
G_R^{\phi^{\dagger}\phi}(x,t)=-i G_R ^{\phi} (x,t)\left[G_F^{\phi}(x,t)-(G_F^{\phi})^{\dagger}(x,t)\right]~.
\end{equation}

Since the poles in $G_F(k,\omega)$ are precisely in its imaginary part, $G_F(x,t)-G_F^*(x,t)$ will have an exponential decay at long times, and therefore our composite operator will thermalize.\footnote{Note that there is also a pole at exactly $\omega=0$. It should be considered trivial: it is simply a result of the fact that a perturbation that is constant in time will not thermalize, which is of course to be expected.}

The response of this operator in a higher dimensional free field theory provides a very good illustration of the physics behind our no-go criterion and the OTH. Diagrammatically we are computing the following one-loop diagram.
\begin{align}
  \label{eq:2}
 \langle {\cal O}(x,t){\cal O}(0,0)\rangle = ~~\raisebox{-.3in}{ 
\begin{tikzpicture}[scale=0.3]
\draw[dashed] (-5,0)--(-3,0);
\draw[thick,decoration={markings, mark=at position 0.25 with {\arrow{>}},  mark=at position 0.75 with {\arrow{>}}},
        postaction={decorate}] (0,0) circle (3);
\draw[dashed] (3,0)--(5,0);
\end{tikzpicture}
}
\end{align}
In higher dimensions it is clear that the operator ${\cal O}(K) = \int \frac{d^dP}{(2\pi)^d}\phi^{\dagger}(P)\phi(K-P)$ sources an essentially infinite set of intermediate sets differentiated by their relative momentum. This large set of intermediate states is then responsible for the decay of the retarded Green's function. OTH is thus a dephasing effect similar to ETH. 
We can show mathematically that this is precisely how this operator fails  our no-go criterion. 
We write
\begin{align}
\phi^\dagger \phi (0,0)=\int \frac{d^{d-1}p}{(2\pi)^{d-1}}\frac{d^{d-1}p'}{(2\pi)^{d-1}} \frac{1}{2\sqrt{\omega(p)\omega(p')}} \left[ a^\dagger_p a_{p'} + a^\dagger_p \bar{a}^\dagger_{p'} + \bar{a}_p a_{p'} + \bar{a}_p \bar{a}^\dagger_{p'} \right]~.
\end{align}
Each of the four terms above has non-zero matrix elements between states that differ by exactly two particles. The energy difference between such states is fixed by the sum (or difference) of the magnitude of the momentum of each of those two particles, but the momentum difference depends on both the magnitudes and the relative orientation of the momenta. For example, consider the first term: 
\begin{equation}
\langle m|a^\dagger_p a_{p'}|n\rangle~.
\end{equation}
This term is non-zero if and only if:\footnote{For simplicity, assume that $\vec{p}\neq \vec{p}'$.}
\begin{equation}
|n\rangle \sim a^\dagger_{p'} a_p|m\rangle~.
\end{equation}
In that case, $\vec{P}_n-\vec{P}_m=\vec{p}'-\vec{p}\equiv \Delta \vec{P}$, and
\begin{align}
E_n-E_m&=\sqrt{\vec{p'}\cdot \vec{p'}+m^2} - \sqrt{\vec{p}\cdot \vec{p}+m^2} \nonumber \\
&=\sqrt{(\Delta \vec{P}+\vec{p})\cdot(\Delta \vec{P}+\vec{p})+m^2}-\sqrt{\vec{p}\cdot\vec{p}+m^2}~.
\end{align}
Since in the operator $:\phi^\dagger \phi (0,0):$ we are instructed to sum over $p$ there is a whole extra vector's worth of information needed to specify the energy difference, even with the momentum difference fixed. This extra vector precisely parametrizes the ``phase-space'' of intermediate states.

To contrast with the mass operator, it is as easy to construct operators that do obey the no-go theorem, even in higher dimensional theories. Consider the following operator: \footnote{This operator is non-local. The connections between non-locality and thermalization are potentially subtle. Nevertheless, we give this example to illustrate explicitly how satisfying the no-go condition prevents the decay of the retarded Green's function}
\begin{align}
\oo(x,t=0)&=\int \frac{d^{d-1}p}{(2\pi)^{d-1}} \left(a_p \bar{a}_p e^{2ip\cdot x} + a^\dagger_p \bar{a}^\dagger_p e^{-2ip\cdot x}\right) \nonumber \\
&= \int \frac{d^{d-1}p}{(2\pi)^{d-1}} \int d^{d-1}y d^{d-1}z \left(\phi(y)\phi(z) e^{ip(y-z+2x)} + \phi^\dagger(y)\phi^\dagger(z) e^{-ip(y-z+2x)}\right) \nonumber \\
&= \int d^{d-1}z\left( \phi(z-2x)\phi(z) + \phi^\dagger(z-2x)\phi^\dagger(z)\right)
\end{align}
It is clear that its matrix elements are non-zero only when the two states involved have $\Delta \vec{P}=\pm 2\vec{p}$ for some $\vec{p}$. In that case, the energy difference between the two states must be $\Delta E=\pm 2\sqrt{\vec{p}\cdot \vec{p} + m^2}=\pm\sqrt{\Delta\vec{p} \cdot \Delta\vec{p} + 4 m^2}$. Our no-go condition is satisfied, and such an operator cannot thermalize.
It is easy to confirm that this operator does not thermalize. Its retarded Green's function is:
\begin{align}
G_R(x,t)&=-i\Theta(t)\langle[\oo(x,t),\oo(0,0)]\rangle_\beta \nonumber\\
&=-i\Theta(t)\int \frac{d^{(d-1)}p}{(2\pi)^{d-1}} \left(e^{2i p \cdot x} - e^{-2 i p\cdot x}\right)\langle a^\dagger_p a_p + \bar{a}^\dagger_p \bar{a}_p \rangle_\beta + \text{(constant)} \nonumber \\
&=-2i\Theta(t) \int \frac{d^{d-1}p}{(2\pi)^{d-1}} \frac{\left(e^{2i p \cdot x} - e^{-2 i p\cdot x}\right)}{e^{\beta \sqrt{\vec{p}^2+m^2}}-1}~,
\end{align}
where on the second line we have dropped the overall infinite additive constant\footnote{Such a constant would contribute only a trivial pole at $\omega=0$.} and used the usual Bose-Einstein distribution. Rather than doing this integral, which has no closed expression in terms of elementary functions, we can simply go to Fourier space and look at the position of the poles to confirm the absence of thermalization. 
\begin{align}
G_R(\vec{k},t)&=-2i\Theta(t) \frac{e^{i\sqrt{\vec{k}^2+4m^2}t}-e^{-i\sqrt{\vec{k}^2+4m^2}t}}{e^{\beta\sqrt{\vec{k}^2+4m^2}}-1} \\
G_R(\vec{k},\omega)&=\frac{2}{e^{\beta\sqrt{\vec{k}^2+4m^2}}-1}\left[\frac{1}{\omega+\sqrt{\vec{k}^2+4m^2}+i\eta}-\frac{1}{\omega-\sqrt{\vec{k}^2+4m^2}+i\eta}\right]~.
\end{align}
We can see that the only poles are at $\omega=\pm\sqrt{\vec{k}^2+4m^2}$, in keeping with our no-go condition.\footnote{Notice that these poles are exactly at the location we expect them to be based on the general arguments given in the Introduction.}

\section{Conclusion}
In this paper, we have demonstrated that integrable and even free field theories have many operators which relax exponentially away into the thermal state of such a system. This is despite the presence of an infinite set of conserved charges. The  conserved charges do constrain the system. We have expressed this constraint on thermalization in terms of a simple no-go condition for the late-time exponential decay of retarded Green's functions. This condition is formulated in terms of the non-zero matrix elements of the operator in question at the origin. We have seen how free fields are restricted from thermalizing because of this condition and how in two dimensions it is powerful enough to stop the thermalization of any chiral operator in conformal field theories, interacting or not. 

We conjecture that this no-go theorem extends directly to integrable theories. Integrable theories are defined by the possibility of expressing any correlation function exactly as a product of two-point correlation functions through Wick's theorem. If it is also possible in such a theory to define a notion of locality in some basis and a pseudo-momentum operator such that pseudo-local operators take the form $\oo(x,t)=e^{iHt-iPx} \oo(0,0) e^{-iHt+iPx}$, then the proof of the no-go theorem carries through.  

More abstractly, for an integrable theory we should be able to write the Hamiltonian as a (weighted) sum of mutually-commuting conserved charges. 
\begin{equation}
H=\sum_i h_i \hat{n}_i~~~,~~~|E\rangle=\otimes_i |n_i\rangle ~~~\text{with}~~~ \hat{n}_i|n_i\rangle=n_i|n_i\rangle~,
\end{equation}
where $i$ labels the different conserved charges, $n_i \in \mathbb{N}$ and $h_i \geq 0$. We can always define an abstract ``pseudo-momentum'' operator with:
\begin{equation}
P=\sum_i p_i \hat{n}_i~,~~p_i \in \mathbb{R}~.
\end{equation}
This operator commutes with the Hamiltonian by construction. We can then define operators that are ``local'' in this basis, and our whole analysis then applies. Of course, whether this pseudo-momentum operator and the associated notion of locality are in any way related to physical space will determine whether this analysis is relevant to observables or not. In the case of free theories, the conserved charges are of course the occupation numbers of the actual momentum modes.

Note that we can always pick the set of coefficients $\{p_i\}$ to be linearly independent from the set $\{h_i\}$. Therefore, given two  energy eigenstates, $\Delta E= \sum_i h_i \Delta n_i$ and $\Delta P= \sum_i p_i \Delta n_i$ are generically independent quantities. A dense operator $\oo$, i.e. one that has generically non-zero matrix elements, has non-zero matrix elements for every point on the $(\Delta E, \Delta P)$ plane. However, an operator satisfying our no-go condition has non-zero matrix elements only when $\Delta E$ is a function of $\Delta P$: this corresponds to a \emph{line} on that $(\Delta E, \Delta P)$ plane: we see that such operators are special (in some sense, sparse). Therefore this tells us that \emph{most} perturbations in an integrable theory violate our no-go condition and should be expected to thermalize. 

We can also compare this with generic interacting field theories. In such theories, states in the Hilbert space are labelled by their total energy and total momentum, and potentially a few other globally conserved charges. Our no-go condition for thermalization still applies, but it is a lot less likely to be satisfied. For example, consider $\lambda \phi^4$ theory with $\phi$ as the operator. The matrix element $\langle m|\phi|n\rangle$ is generically non-zero because $|m\rangle$ and $|n\rangle$ cannot be thought of as states built out of a definite number of particles. From the point of view of the calculation of the retarded Green's function, the time evolution of $\phi$ is much more complicated than in the free theory. In particular, $\phi(x,t)=e^{iHt}\phi(x,0)e^{-iHt}$ is not linear in creation and annihilation operators defined at $t=0$, therefore the commutator $[\phi^\dagger(x,t),\phi(0,0)]$ is not a c-number. Of course, if one were to pick a fine-tuned operator to satisfy our no-go condition, we would still expect it not to thermalize. 

This discussion raises the distinction of thermalization due to a specific choice of operator vs thermalization due to dynamics (vs eigenstate thermalization). For free field theories, we have shown how a suitable choice of operator leads to thermalization of perturbations. But that thermalization was in some sense caused by our choice of operator and state. Specifically, we started with a thermal state and then perturbed it with an operator that was spread over enough of the phase space to dissipate. The natural Hamiltonian evolution of the system did little to actually facilitate the exploration of the phase space. The "size" of the operator, in the sense of how many modes it couples, is independent of time. 
In contrast, in an interacting theory, as operators evolve in time they start coupling more and more modes. Perturbatively, we have illustrated above how we can understand this in terms of Feynman diagrams. In a free theory, there are no interaction vertices and we can only draw straight lines (i.e. Feynman propagators). The retarded Green's function can be expressed as a difference of amplitudes. The way we get thermalization is by using \emph{composite} operators: this is equivalent to manually inserting operators that join into vertices. By contrast, in an interacting theory it is the Hamiltonian itself that provides the vertices as part of the Feynman rules. 

Qualitatively this operator thermalization shares two properties with eigenstate thermalization\footnote{for recent connections between ETH and CFT methods, see for instance \cite{Dymarsky:2016ntg,Brehm:2018ipf,Romero-Bermudez:2018dim, Hikida:2018khg,Anous:2019yku,Nayak:2019khe}.}: the thermalization is not driven by dynamics, and an initial condition spreads out over a large part of the system. On the other hand, there are distinct differences as well. Eigenstate thermalization happens in a generic closed quantum system with few to no conserved charges for a small distinct set of operators. Increasing the set of operators reveals the original pure state. Operator thermalization happens generically but becomes important for a small set of distinct theories that have an infinite set of conserved charges. 

Coming back to the previous point, it would be interesting to explore this difference between operator and Hamiltonian thermalization in more detail. An obvious starting point is to see how our story can be adapted to out-of-time-order correlators (OTOCs), as opposed to retarded Green's functions. OTOCs or refinements thereof are supposed to measure operator growth \cite{Lucas:2018wsc,Qi:2018bje}. This operator growth is the quantum analogue of chaotic behavior underlying the ergodic theorem. In other words, the intractability of operator growth to a unique initial condition is responsible for entropy growth, dissipation and hence thermalization. It therefore appears integral to Hamiltonian thermalization. Nevertheless one would surmise from our Operator Thermalization Hypothesis that even OTOCs in free theories can show similar ``mixing'' behavior. 

Finally, we should move beyond from linear response, to consider other states as initial configuration, and to connect to the many new insights in non-equilibrium phenomena in integrable theories discovered in recent years. We refer in particular to the highly active research effort into the formation of the Generalized Gibbs Ensemble in 2D integrable systems after a quench 
\cite{2007PhRvL..98e0405R,Caux:2013ra,2015PhRvL.115o7201I,2016JSMTE..06.4006C,2016JSMTE..06.4007V,2016JSMTE..06.4002E,2017PhRvB..95k5128I} and the associated dissipation to reach this state
\cite{2016JSMTE..06.4007V,2016JSMTE..06.4002E,Castro-Alvaredo:2016cdj,Bertini:2016tmj,2017arXiv171100873C,2017PhRvB..95k5128I,2018PhRvL.120d5301D,DeNardis:2018omc}. Our study has shown that conformal invariance very strongly constrains thermalization: chiral operators can never thermalize, no matter the interactions and couplings in the theory. It is known that 2D CFTs have an infinite tower of conserved charges, related to the KdV hierarchy. It would be interesting to see if our linear response statements dictating which operators thermalize and which do not are related to the onset of the GGE. The thermalization we have discussed in this paper is certainly a necessary condition for the evolution of pure states towards states accurately described by a Gibbs ensemble, but it is not sufficient. 

A strong hint that our no-go criterion might be relevant more generally is the following. So far, we have discussed the response of an operator to itself, i.e. the pole structure of the commutator of an operator with itself. But our analysis can easily be extended to study the response of any other operator to a perturbation of a given operator. Physically, we mean the following: build a thermal state of an unperturbed Hamiltonian and then introduce a perturbation by coupling an operator $\oo_1$ to a classical source. The response of any other observable $\oo_2$ will be given by the commutator $[\oo_2(x,t),\oo_1(0,0)]$. Using the exact same arguments as before and noticing that $\langle m|\oo_1|n\rangle=0$ if and only if $|\langle m|\oo_1|n\rangle|^2=0$, we can see that if $\oo_1$ obeys the no-go condition, then \emph{no observable} will lose track of that perturbation. Conversely, if $\oo_1$ violates the no-go condition, then we expect that the response of most observables (the exception being the observables that themselves obey the condition) will relax away. This extends our analysis from the response of a single observable to the response of most observables, which gives more insight into the structure of the state itself.

\acknowledgments
The authors have benefited from discussions with Aleksandar Bukva, Jean-S\'{e}bastian Caux, Vadim Cheianov, Sasha Krikun, Neil Robinson, Aurelio Romero-Bermudez, Vincenzo Scopelliti, and especially Enej Ilievski.
This research was supported in part by a VICI award of the Netherlands Organization for Scientific Research (NWO), by the Netherlands Organization for Scientific Research/Ministry of Science and Education (NWO/OCW), by the Foundation for Research into Fundamental Matter (FOM) and by the National Science and Engineering Research Council of Canada (NSERC). PSG thanks the University of British Columbia's Physics and Astronomy department for hospitality during part of this work.

\appendix
\section{Details of vertex operator calculation}
\begin{align}
\langle:e^{A_1+A_2}:\rangle_\beta&=\frac{1}{Z} \text{Tr}\left[e^{-\beta H} \prod_{k=1}^\infty \left(e^{-\gamma_k^* a_k^\dagger-\bar{\gamma}_{k}^*a_{-k}^\dagger}\right)\prod_{k=1}^\infty \left( e^{\gamma_k a_k+\bar{\gamma}_k a_{-k}}\right)  \right] \\
&=\left[\prod_{k=1}^\infty \sum_{n_k=0}^\infty \frac{e^{-\beta E_k}}{Z_k} \langle n_k|e^{-\gamma_k^* a_k^\dagger} e^{\gamma_k a_k}|n_k\rangle\right]\left[\prod_{k=1}^\infty \sum_{n_k=0}^\infty \frac{e^{-\beta E_k}}{Z_k} \langle n_{-k}|e^{-\bar{\gamma}_k^* a_{-k}^\dagger} e^{\bar{\gamma}_k a_{-k}}|n_{-k}\rangle\right]~, 
\end{align}
where $E_k$ and $Z_k$ are the energy and partition function of a simple harmonic oscillator of frequency $\omega_k$.
The Baker-Campbell-Hausdorff formula gives us that
\begin{equation}
e^{-\gamma_k^* a^\dagger_k} e^{\gamma_k a_k}=e^{|\gamma_k|^2} e^{\gamma_k a_k} e^{-\gamma_k^* a^\dagger_k}~,
\end{equation}
so that
\begin{align}
\langle n_k|e^{-\gamma_k^* a_k^\dagger} e^{\gamma_k a_k}|n_k\rangle&=e^{|\gamma_k|^2} \langle 0_k|e^{\gamma_k a_k} \frac{(a_k)^{n_k}}{\sqrt{n_k!}}\frac{(a_k^{\dagger})^{n_k}}{\sqrt{n_k!}} e^{-\gamma_k^* a_k^\dagger}|0_k\rangle \\
&=e^{2|\gamma_k|^2} \langle \gamma_k^*|\frac{(a_k)^{n_k}}{\sqrt{n_k!}}\frac{(a_k^{\dagger})^{n_k}}{\sqrt{n_k!}}|-\gamma_k^*\rangle \\
&=\sum_{m_k,m'_k} e^{|\gamma_k|^2} \langle m_k |\frac{(\gamma_k)^{m_k}}{\sqrt{m_k!}} \frac{(a_k)^{n_k}}{\sqrt{n_k!}}\frac{(a_k^{\dagger})^{n_k}}{\sqrt{n_k!}} \frac{(-\gamma_k^*)^{m'_k}}{\sqrt{m'_k!}}|m'_k\rangle~\\
&=e^{|\gamma_k|^2}\sum_{m_k}^\infty (-|\gamma_k|^2)^{m_k} \frac{(m_k+n_k)!}{m_k!} \frac{1}{m_k! n_k!}
\end{align}
where we have used the usual coherent states $|\gamma_k\rangle=e^{-|\gamma_k|^2/2}e^{\gamma_k a^\dagger_k}|0_k\rangle$~.
Therefore, 
\begin{align}
\sum_{n_k=0}^\infty \frac{e^{-\beta E_k}}{Z_k} \langle n_k|e^{-\gamma_k^* a_k^\dagger} e^{\gamma_k a_k}|n_k\rangle&=\frac{e^{|\gamma_k|^2}}{\sum_{n_k=0}^\infty e^{-\beta n_k \omega_k}}\sum_{m_k,n_k=0}^\infty  e^{-\beta n_k \omega_k} (-|\gamma_k|^2)^{m_k} \frac{(m_k+n_k)!}{(m_k!)^2 n_k}\\
&=e^{-\frac{|\gamma_k|^2}{e^{\beta \omega_k}-1}}~.
\end{align}
We have
\begin{align}
\langle:e^{-i\phi(0,t)+i\phi(0,0)}:\rangle_\beta&= \exp\left(-\frac{|\gamma_0|^2}{e^{\beta \omega_0}-1}\right)\prod_{k=1}^\infty \exp\left(-\frac{|\gamma_k|^2}{e^{\beta \omega_k}-1}\right)\exp\left(-\frac{|\bar{\gamma}_k|^2}{e^{\beta \omega_k}-1}\right)\nonumber \\
&=\exp\left(-\sum_{k=1}^\infty \frac{|\gamma_k|^2+|\bar{\gamma}_k|^2}{e^{\beta \omega_k}-1}-\frac{|\gamma_0|^2}{e^{\beta \omega_0}-1}\right)
\end{align}

\begin{equation}
|\gamma_k|^2=\frac{\xi^2}{2\pi k} \left[1-\cos\left(\frac{2\pi}{L} k(t-x)\right)\right]~~,~~|\bar{\gamma}_k|^2=\frac{\xi^2}{2\pi k} \left[1-\cos\left(\frac{2\pi}{L} k(t+x)\right)\right]
\end{equation}

A useful fact is:
\begin{equation}
\frac{1}{e^{\beta x}-1} = \sum_{m=1}^\infty e^{-\beta x m}~.
\end{equation}
We have
\begin{align}
\xi^2\sum_{k=1}^\infty \frac{1-\cos\left(\frac{2\pi k}{L} (t-x)\right)}{2\pi k\left(e^{2\pi k\beta/L}-1\right)}=\xi^2\sum_{m=1}^\infty \sum_{k=1}^\infty \frac{1-\cos\left(\frac{2\pi k}{L} (t-x)\right)}{2\pi k} e^{-2\pi k m \beta/L} \nonumber\\
=\frac{\xi^2}{4\pi}\sum_{m=1}^\infty\left[\log\left(1-e^{\frac{2\pi i(t-x+im\beta)}{L}}\right)+\log\left(1-e^{\frac{-2\pi i(t-x-im\beta)}{L}}\right)-2\log\left(1-e^{-\frac{2\pi m\beta}{L}}\right)\right]
\end{align}
Therefore,
\begin{align}
\exp\left[-\sum_{k=1}^\infty \frac{|\gamma_k|^2}{e^{\beta \omega_k}-1}\right]&=\prod_{m=1}^\infty\left[\frac{\left(1-e^{-2\pi m \beta/L}\right)^2}{\left(1-e^{\frac{2\pi i}{L}(t-x+im\beta)}\right)\left(1-e^{\frac{-2\pi i}{L}(t-x-im\beta)}\right)}\right]^{\frac{\xi^2}{4\pi}}\\
&=\prod_{m=1}^\infty\left[\frac{2\sinh^2\left(\frac{\pi m \beta}{L}\right)}{\cosh\left(\frac{2\pi m \beta}{L}\right)-\cos\left(\frac{2\pi (t-x)}{L}\right)}\right]^{\frac{\xi^2}{4\pi}}
\end{align}
Expanding the factors in powers of $1/L$, we have
\begin{align}
\exp\left[-\sum_{k=1}^\infty \frac{|\gamma_k|^2}{e^{\beta \omega_k}-1}\right]&=\prod_{m=1}^\infty\left[\frac{m^2\beta^2}{(t-x)^2+m^2\beta^2} +\mathcal{O}(1/L^2) \right]^{\frac{\xi^2}{4\pi}} \\
&=\left(\frac{\pi (t-x)}{\beta} \frac{1}{\sinh\left(\frac{\pi (t-x)}{\beta}\right)}\right)^{\frac{\xi^2}{4\pi}}~,
\end{align}
where in the last line we have taken the $L\rightarrow\infty$ limit to recover the result on the line. 
Similarly,
\begin{equation}
\exp\left[-\sum_{k=1}^\infty \frac{|\bar{\gamma}_k|^2}{e^{\beta \omega_k}-1}\right]=\left(\frac{\pi (t+x)}{\beta} \frac{1}{\sinh\left(\frac{\pi (t+x)}{\beta}\right)}\right)^{\frac{\xi^2}{4\pi}}
\end{equation}

Putting everything together, we get
\begin{equation}
\langle :e^{-i\xi\phi(x,t)+i\xi\phi(0,0)}: \rangle_\beta \sim \left(\frac{\pi(t+x)}{\beta\sinh\left(\frac{\pi (t+x)}{\beta}\right)}\right)^{\frac{\xi^2}{4\pi}} \left(\frac{\pi(t-x)}{\beta\sinh\left(\frac{\pi (t-x)}{\beta}\right)}\right)^{\frac{\xi^2}{4\pi}}
\end{equation}

\bibliographystyle{jhep}
\bibliography{tfi}

\end{document}